\documentclass[aps,prb,a4paper,notitlepage,onecolumn,noshowpacs,longbibliography,10pt]{revtex4-1}

\usepackage{graphicx}
\usepackage[dvipsnames]{xcolor}
\usepackage{amsmath}
\usepackage{siunitx}

\begin{document}

\title{Nonlinear two-level dynamics of quantum time crystals}

\author{S.~Autti$^{1,2\ast}$}

\author{P.J. Heikkinen$^{1,3}$}

\author{J. Nissinen$^{1}$}

\author{J.T. M\"akinen$^{1}$}

\author{G.E.~Volovik$^{1,4}$}

\author{V.V. Zavyalov$^{1,2}$}

\author{V.B.~Eltsov$^1$}

\vspace{0.4cm}
\affiliation{$^1$Low Temperature Laboratory, Department of Applied Physics, Aalto University, POB 15100, FI-00076 AALTO, Finland.\\
$^2$Department of Physics, Lancaster University, Lancaster LA1 4YB, UK *Email: s.autti@lancaster.ac.uk\\
$^3$Department of Physics, Royal Holloway University of London, Egham, Surrey, TW20 0EX, UK.\\
$^4$L.D. Landau Institute for Theoretical Physics, Moscow, Russia}
\maketitle


{\bf A time crystal is a macroscopic quantum system in periodic motion in its ground state \cite{PhysRevLett.109.160401,PhysRevLett.111.070402,Sacha_2017,PhysRevLett.120.215301}, stable only if isolated from energy exchange with the environment. For this reason, coupling separate time crystals is challenging\cite{autti2020}, and time crystals in a dynamic environment have yet not been studied. In our experiments, two coupled time crystals made of spin-wave quasiparticles (magnons) form a macroscopic two-level system. The two levels evolve in time as determined intrinsically by a nonlinear feedback \cite{Autti2012}. Magnons move from the ground level to the excited level driven by the Landau-Zener effect, combined with Rabi population oscillations. We thus demonstrate how to arrange spontaneous dynamics between interacting time crystals. Our experiments allow access to every aspect and detail of the interaction in a single run of the experiment, inviting technological exploitation\cite{chumak2015magnon} -- potentially even at room temperature\cite{alex2019josephson,mohseni2020bose,PhysRevLett.126.057201}.  
}

Perpetual periodic motion in an equilibrium state that can be observed is famously unfeasible: in experiments, time crystals bend either the  equilibrium\cite{zhang2017observation,choi2017observation,rovny2018observation} or the perpetuity\cite{PhysRevLett.120.215301} requirement. Spin systems have proved the best platform to balance this delicate combination\cite{Sacha_2017,else2020discrete}. A common way to create a time crystal is by applying an external drive continuously\cite{Sacha_2017,else2020discrete}, but this may interfere with the dynamics in the system. In the superfluid B phase of $^3$He ($^3$He-B), magnon time crystal life time can be extended up to a thousand seconds\cite{2000_ppd}, approximating equilibrium in the absence of pumping. In this Letter, we make use of this advantage to study the spontaneous dynamics of two magnon time crystals trapped within $^3$He-B~\cite{autti2020}. The chemical potentials of the time crystals, or the energy levels in the trap, cross spontaneously. We show that the resulting dynamics follow the textbook description of a two-level system, modified by the nonlinear feedback. 

Magnon time crystals in superfluid $^3$He consist of the quanta of transverse spin waves. Spin waves are associated with magnetisation that precesses about the external magnetic field $\mathbf{H}$. At sufficient magnon density and low enough temperature, Bose-Einstein condensation of magnons leads to the transition to the time crystal
phase: the precession synchronises spontaneously at uniform frequency $\omega$ manifesting the characteristic periodic motion in the time crystal\cite{PhysRevLett.120.215301,autti2020}. 

\begin{figure}[hbt!]
 \centerline{\includegraphics[width=1\linewidth]{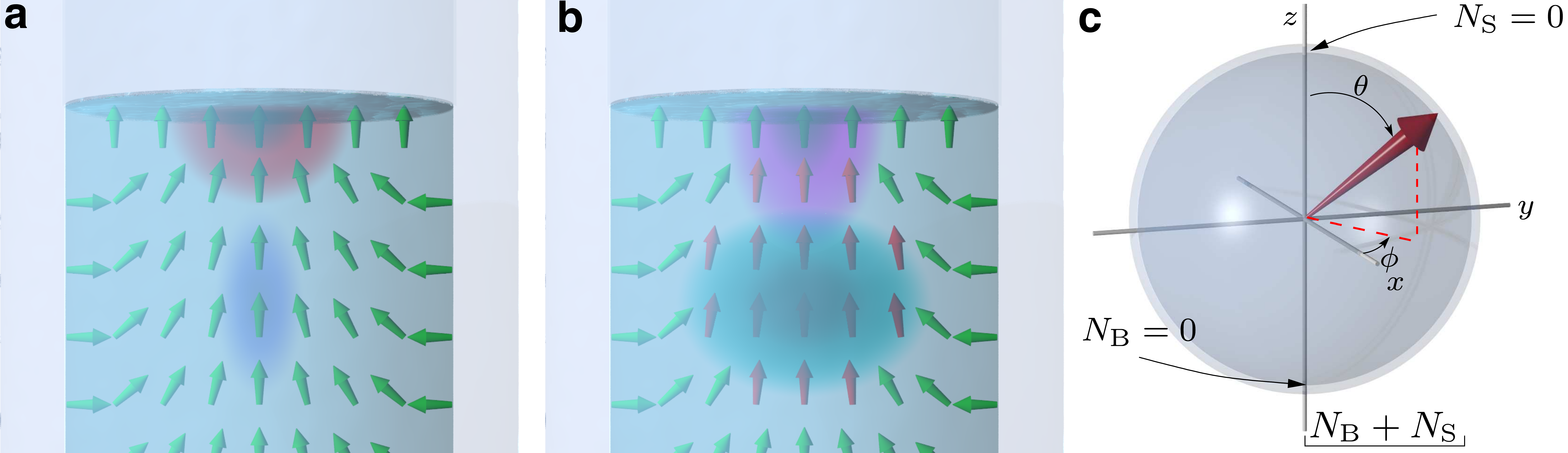}}
   \caption{\label{Fig:coupling_schematic} {\bf Time crystal two-level system:} ({\bf a}) The superfluid sample is contained in a glass cylinder (vertical walls) and is limited from above by a free surface. The resulting distribution of $\mathbf{L}$ (green arrows) confines magnons in two local minima, hosting two adjacent time crystals: one in the bulk of the superfluid (blue blob) and the other one touching the free surface (red blob). In each time crystal, magnetisation is precessing coherently, which couples to measurement circuitry as shown in Extended Data Fig.~\ref{Fig:experiment}. ({\bf b})  Magnons in the bulk modify the confining trap created by the $\mathbf{L}$ distribution. When the bulk population is large (cyan blob), the textural trap is widened (red arrows), which modifies also surface time crystal's wave function (magenta blob). This increases the coupling between the states. Changes in the trap and the wave functions have been exaggerated for illustrational purposes. ({\bf c}) The state of the two-level system (red arrow) can be illustrated using a Bloch sphere where the sphere surface corresponds to magnon number $N_\mathrm{B}+N_\mathrm{S}$, and the interior to smaller magnon numbers. The relative phase between the time crystals' precession corresponds to the the azimuthal angle $\phi$, and the polar angle $\theta$ describes the population distribution (see Methods). }
 \end{figure}

We use the spin-orbit interaction to create a trap for two adjacent time crystals: In the sample container cylinder, the macroscopic distribution of the average orbital momentum $\mathbf{L}$ is axially symmetric (green arrows in Fig.~\ref{Fig:coupling_schematic}a), and magnons are trapped by that distribution. The trap is fine-tuned using a magnetic field profile as detailed in Methods. The resulting harmonic trap has a minimum in the middle of the sample container. A second local minimum, separated from the bulk minimum by a constriction, is created at the adjacent free surface of the superfluid\cite{autti2020}. In what follows, we concentrate on the lowest level in each trap and measure all frequencies in the frame rotating at the Larmor frequency $\omega_0\propto |\mathbf{H}|$ taken at the centre of the bulk trap.

The combination of the two time crystals in local minima of the potential creates a two-level system, conveniently illustrated by a macroscopic Bloch sphere (Fig.~\ref{Fig:coupling_schematic}c). Let us denote the bulk time crystal population $N_\mathrm{B}$ and the surface population $N_\mathrm{S}$. The bulk and surface precession frequencies ($\omega_\mathrm{B}$, $\omega_\mathrm{S}$) and the coupling $\Omega$ between the crystals are determined by the profile of the confining trap, as detailed in Methods. The dynamics of the coupled levels are described by the Hamiltonian 

\begin{gather}\label{Hamiltonian}
 \mathcal{H}=\hbar
 \begin{pmatrix} \omega_\mathrm{B}[N_\mathrm{B}(t)] & -\Omega \\ -\Omega & \omega_\mathrm{S} \end{pmatrix},
\end{gather}
where $\hbar$ is the reduced Planck constant, $t$ is time, and the role and origin of the dependence $\omega_\mathrm{B}[N_\mathrm{B}(t)]$ is detailed below.  

\begin{figure}[hbt!]
 \centerline{\includegraphics[width=1.0\linewidth]{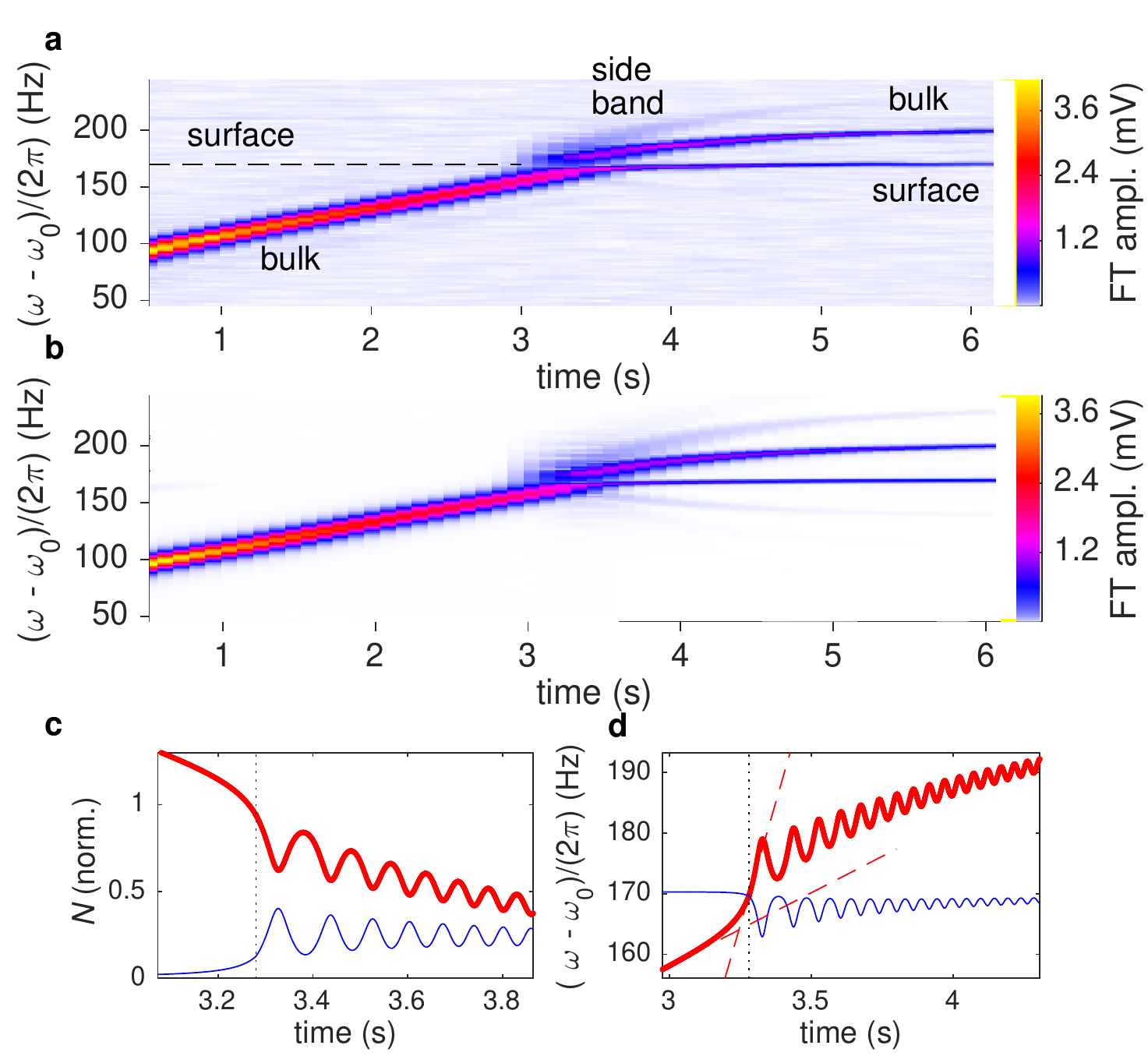}}
   \caption{\label{Fig:splitting2} {\bf Two-level time crystal dynamics:} ({\bf a}) The signal from the pick-up coils, analysed with windowed Fourier transformation (FT), shows the bulk time crystal, populated at $t=0$, as a moving sharp peak. Initially $\omega_\mathrm{B}<\omega_\mathrm{S}$, but as the population in the bulk trap decays, at $t=3.3$~s, the global ground state moves to the surface in an avoided crossing. The excited state, now located in the bulk, is simultaneously populated. Rabi (Josephson) population oscillations are seen as a side band. Coupling extracted from the side band extrapolates to $\Omega/(2 \pi)\approx1$~Hz at the crossing, in good agreement with the fitted simulation value $\Omega/(2 \pi)\approx1.4$~Hz.({\bf b}) The numerical simulation recreates the population transfer and the side band, ({\bf c}) showing that at the crossing the population is distributed between the bulk trap (thick red line) and the surface trap (thin blue line) in a Landau-Zener transition, followed by population oscillations. Total population is normalised to one at the crossing. ({\bf d}) Simulated $\omega_\mathrm{B}$ (thick red line) and $\omega_\mathrm{S}$ (thin blue line), cross at $t=3.3$~s (dotted vertical line). Corresponding dressed frequencies are shown in Extended Data Fig.~\ref{Fig:splitting3}. Dashed red lines illustrate the increase in $\mathrm{d} |\omega_\mathrm{S}-\omega_\mathrm{B}|/\mathrm{d}t\approx \mathrm{d} \omega_\mathrm{B}/\mathrm{d}t$. In this measurement temperature was \SI{180}{\micro\kelvin} and $\omega_0/(2 \pi)=833$~kHz. 
   }
 \end{figure}

In experiments, the levels can be populated in desired proportion by a radio-frequency pulse via adjacent coils. To highlight the two-level dynamics, we populate only the bulk time crystal in the beginning of the experiment shown in Fig.~\ref{Fig:splitting2}a. After the pulse, the coherent precession of magnetisation induces an oscillating signal in the coils, which allows inferring the frequency and phase, and the signal amplitude yields the magnon number. The pumping is followed by exponential decay of $N_\mathrm{B}$ with time constant $\tau_\mathrm{B}$, controlled by temperature as detailed in Methods.   

The spontaneous evolution of $\omega_\mathrm{B}$ seen in the experiment is created by the underlying superfluid system. Due to spin-orbit coupling, the bulk trap is widened for increasing $N_\mathrm{B}$, resulting in decreasing $\omega_\mathrm{B}$\cite{Autti2012}. The effect is schematically illustrated in Fig.\ref{Fig:coupling_schematic}b. In Fig.~\ref{Fig:splitting2}, the ground level is initially located in the bulk trap, $N_\mathrm{B}$ decays at a rate determined by $\tau_\mathrm{B}$, and $\omega_\mathrm{B}$ increases slowly as the trap recovers a narrower shape. The boundary conditions fix the shape of the surface trap ($\omega_\mathrm{S}$ is constant), and $\omega_\mathrm{B} > \omega_\mathrm{S}$ when $N_\mathrm{B}$ is small. Hence, $\omega_\mathrm{B}$ eventually crosses $\omega_\mathrm{S}$ before levelling out. 

In a coupled two-level system, a level crossing has specific consequences: The observed frequencies are the (dressed) eigenfrequencies of the Hamiltonian which deviate from the undressed frequencies $\omega_\mathrm{B},\omega_\mathrm{S}$ in the Rabi regime $\Omega> |\omega_\mathrm{B}-\omega_\mathrm{S}|$. Due to this hybridisation, the observed levels avoid crossing each other, and the global ground level smoothly switches from bulk to surface (from $\omega_\mathrm{B}$ to $\omega_\mathrm{S}$) as seen in Fig.~\ref{Fig:splitting2}a. Population transfer between the levels is also observed, as both of the levels are populated after the avoided crossing. 

Traversing the avoided crossing adiabatically (slowly enough) would allow the entire magnon population to follow the global ground state, but here some magnons move to the state with higher eigenenergy (precession frequency). This process is generally known as Landau-Zener-Stueckelberg-Majorana tunnelling, below referred to as Landau-Zener tunnelling. The population transferred depends on the rate of level crossing  $\mathrm{d} |\omega_\mathrm{S}-\omega_\mathrm{B}|/\mathrm{d}t$ at $\omega_\mathrm{B}=\omega_\mathrm{S}$. For usual coupled non-linear oscillators with damping, the crossing rate is determined directly by the damping. Here the decay rate $\tau_\mathrm{B}$, inserted into $\omega_\mathrm{B}[N_\mathrm{B}(t)]$, corresponds to population transfer two orders of magnitude smaller than that observed in the experiment. We can analyse this striking mismatch by simulating the time evolution of the two-level Hamiltonian numerically (Methods). 

The simulation yields quantitative agreement with the experiment (Fig~\ref{Fig:splitting2}b). We can therefore use the simulation to extract the populations (Fig~\ref{Fig:splitting2}c) and the undressed frequencies (Fig~\ref{Fig:splitting2}d) needed to analyse the Landau-Zener tunnelling. Near the avoided crossing ($\Omega > |\omega_\mathrm{B}-\omega_\mathrm{S}|$) the two time crystals exchange population via Rabi oscillations. Rabi oscillations in a flexible trap increase the population transfer probability by accelerating the crossing rate (Fig~\ref{Fig:splitting2}d). Importantly, this effect is caused by the flexible trap being modified by the population oscillation, not by level hybridisation. Inserting the accelerated crossing rate found in the simulation to the Landau-Zener formula (see Methods) yields the predicted population transfer of 65\%, in good agreement with the simulated population transfer, 63\%. That is, the instantaneous crossing rate at the avoided crossing determines the population transfer magnitude even when the rate of frequency change is rapidly varying near the crossing. Hence, we confirm that the Landau-Zener description is valid even if the crossing rate is regulated by intrinsic feedback. 

Far from the avoided crossing, the two-level interaction is characterised by AC Josephson population oscillations between the levels\cite{autti2020}. Owing to the feedback in the bulk trap, the oscillations result in a side band that follows the bulk trace, separated from the main trace by the Josephson frequency $|\omega_\mathrm{B}-\omega_\mathrm{S}|$ (Methods). Put together, the population oscillations and the population transfer confirm that the two time crystals form a macroscopic two-level system.

\begin{figure}[hbt!]
 \centerline{\includegraphics[width=1\linewidth]{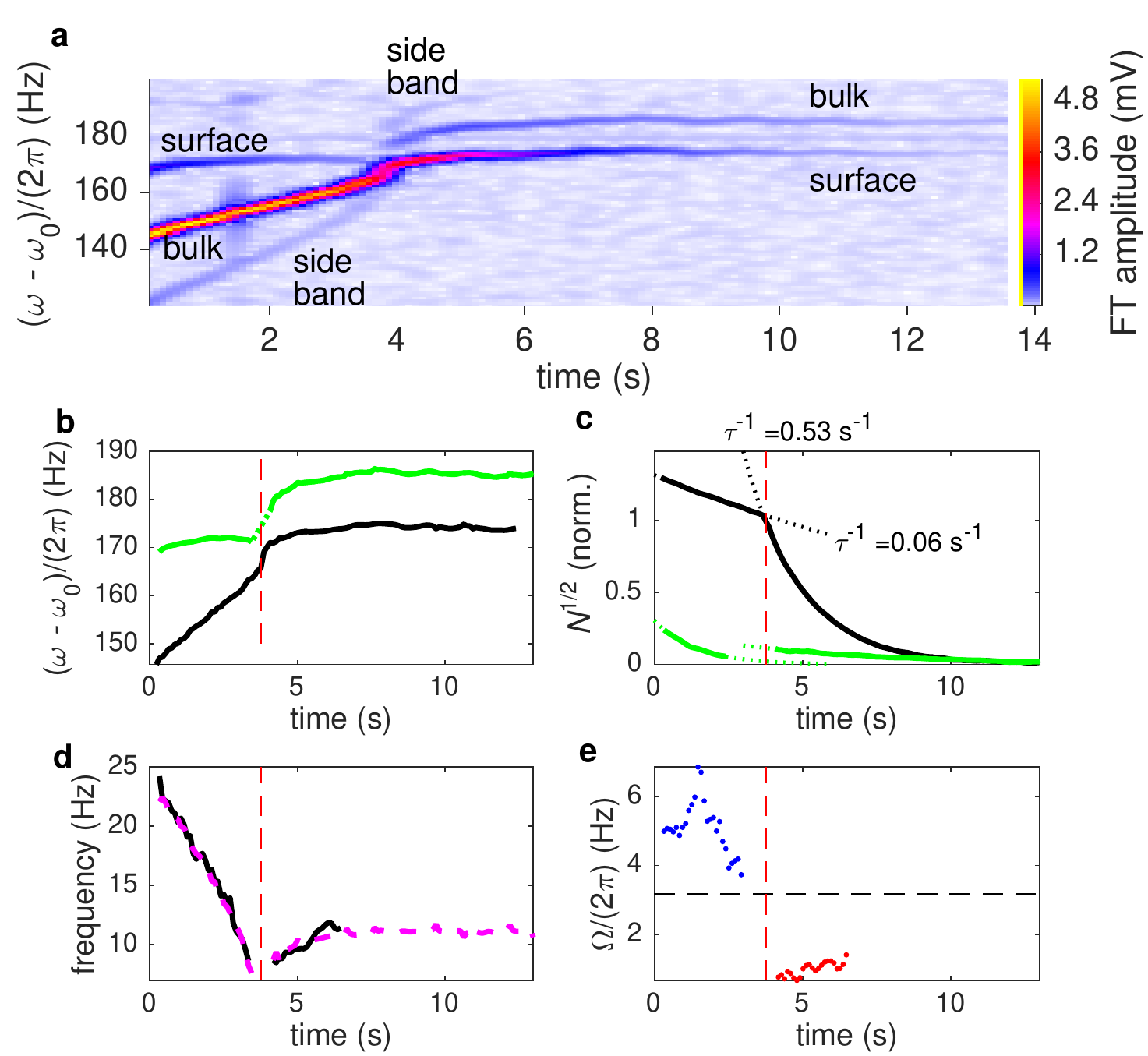}}
   \caption{\label{Fig:splitting} {\bf Two-level time crystal with dynamic coupling:} ({\bf a}) The time crystals are created at $t=0$. ({\bf b}) Initially the ground level (black line) is located in the bulk and the excited level (solid green line) at the surface. At $t\approx3.8$~s (vertical dash lines), the ground level moves smoothly to the surface in an avoided crossing. Dotted green line shows a linear interpolation of the excited-state frequency at the crossing. ({\bf c}) Most of the population follows the ground level (black line) movement from bulk to surface, identified by a sharp increase in the exponential relaxation rate (fitted dash lines). Total population is normalised to one at the crossing. ({\bf d}) Population oscillations between the time crystals are seen as a side-band of the bulk crystal trace in panel {\bf a}. The side band's frequency separation from the bulk trace (solid black line) is equal to the frequency separation of the main traces (magenta dash line). ({\bf e}) The coupling $\Omega$ can be extracted from the side band and main trace amplitudes, in good agreement with that estimated by linear interpolation from the separation of the main traces in panel {\bf b} (horizontal dash line). In this measurement temperature was \SI{150}{\micro\kelvin} and $\omega_0/(2 \pi)=624$~kHz.
   }
 \end{figure}

It is worth noting that magnon time crystal dynamics, enhanced by the nonlinear feedback, can be analysed directly from the experiments without resorting to a numerical simulation of the system. This unique advantage will allow untangling interactions involving multiple time crystals that go beyond the two-level description. As a simple demonstration of this capability, we introduce a level crossing in a region where $N_\mathrm{B}$ is large. Both levels are populated in the beginning of the experiment (Fig.~\ref{Fig:splitting}a) to allow following their dynamics directly. 

In this experiment the coupling is changing, as detailed below, and the system no longer obeys the Hamiltonian (\ref{Hamiltonian}). Otherwise the dynamics follow a similar pattern: The ground state moves from bulk to surface (Fig.~\ref{Fig:splitting}b), and the moment when this happens is identified by a sharp increase in the observed relaxation rate (Fig.~\ref{Fig:splitting}c). The increase is attributed to increased dissipation in the surface trap due to surface-mediated emission of other spin wave modes\cite{HiggsNComm} and potentially surface-bound Majorana states\cite{Muarakawa2011,chung_prl103}. Josephson population oscillations are seen as the side band, separated from the bulk trace by $|\omega_\mathrm{B}-\omega_\mathrm{S}|$ (Fig.~\ref{Fig:splitting}d). The coupling, extracted from the amplitude of the population oscillations (Fig.~\ref{Fig:splitting}e), is the largest in the beginning of the experiment and decreases when $N_\mathrm{B}$ decreases. That is, the constriction between the time crystals is affected by the bulk trap modification (Fig.~\ref{Fig:coupling_schematic}b), which makes the coupling larger when $N_\mathrm{B}$ is large.

In conclusion, we show that the dynamics and interactions of the two adjacent magnon time crystals are quantitatively described by a two-level Hamiltonian. The levels are modified by a nonlinear feedback, arising owing to spin-orbit interaction in the underlying superfluid system. This allows engineering intrinsic time crystal dynamics, reducing decoherence and noise by removing the need for continuous external drive. We show that when the two-level eigenfrequencies approach one another, the coupling between the levels results in an avoided crossing with ensuing Landau-Zener population transfer from the global ground state to the excited state and Rabi population oscillations. The simulated population dynamics agree with the experiment, confirming the validity of the two-level Hamiltonian. All relevant observables and parameters including the eigenfrequencies and the coupling between the time crystals can be simultaneously extracted from the experiment. We emphasise that each measurement sequence presented corresponds to a single run of the experiment, and the phenomena are well reproducible. 

We have shown that the spin-orbit interaction can be harnessed to create a nonlinear feedback for magnons in a coherent time crystal system. Nonlinear feedback is needed for spin-based versions of quantum devices such as the SQUID. We emphasise that Bose-Einstein condensates of magnons with similar
coherent dynamics are accessible also in a solid-state room temperature system\cite{Bozhko20161057,alex2019josephson,BEC_Josephson,PhysRevLett.126.057201}, promising on-chip applications in ambient conditions. It remains an interesting task to demonstrate parametric pumping of magnons and logic gate operations between the two levels using an additional magnetic trap that can be manipulated externally. Any number of co-existing time crystals can be accommodated in a magnetic landscape to increase the number of degrees of freedom, and the flexible trap can be turned off by adjusting the external magnetic field. These are important steps for realising magnon-based devices and information processing\cite{byrnes2012macroscopic,andrianov2014magnon,chumak2015magnon,lachance2019hybrid,tabuchi2016quantum,clerk2020hybrid}. To access phenomena such as quantum entanglement, few-magnon operations can be implemented using nano-fluidic confinement and ultra-sensitive NMR techniques\cite{Levitin841,heikkinen2020fragility}. Finally, the hybridised two-level spectrum makes an extremely sensitive probe for the dynamics in the two-dimensional system of surface-bound quasiparticles in superfluid $^3$He~~\cite{autti2020fundamental,lotnyk2020thermal}, including elusive surface-bound Majorana fermions that are expected to manifest themselves as zero-temperature magnetic dissipation \cite{Muarakawa2011,chung_prl103}. 

\section*{Acknowledgements}

We thank A. Veps\"{a}l\"{a}inen for stimulating discussions. This work has been supported by the European Union's Horizon 2020 research and innovation programme (694248). The experimental work was carried out in the Low Temperature Laboratory, which is a part of the OtaNano research infrastructure of Aalto University and of the EU H2020 European Microkelvin Platform (824109). S.A. and V.V.Z. were funded by UK EPSRC (EP/P024203/1). S.A. acknowledges support from the Jenny and Antti Wihuri foundation, and P.J.H. that from the V\"{a}is\"{a}l\"{a} foundation of the Finnish Academy of Science and Letters. We acknowledge the computational resources provided by the Aalto Science-IT project.
	
\section*{Competing interests}
The authors declare no competing interests.

\section*{Author contributions}
The manuscript was written by S.A. with contributions from all authors. Experiments were planned, carried out, and analysed by S.A., J.T.M, P.J.H, V.V.Z, and V.B.E. Theoretical work was done by S.A., G.E.V., J.N. and V.B.E. The project was supervised by V.B.E.

\bibliography{TimeCrystal_LZ_v3.9_arxiv}
\bibliographystyle{naturemag}

\section*{Methods}
\subsection*{Experiment}

The superfluid $^3$He sample is placed in a cylindrical quartz-glass container (15~cm long, 6~mm diameter) in a nuclear demagnetisation refrigerator (Extended Data Fig.~\ref{Fig:experiment}). The lower end of the sample container connects to a volume of sintered silver powder surfaces, thermally connected to the nuclear refrigerant. This allows cooling the $^3$He down to \SI{130}{\micro\kelvin}. Temperature of the superfluid is measured using a quartz tuning fork\cite{2007_forks,2008_forks}, and pressure is equal to saturated vapour pressure, which is near zero at these low temperatures. The superfluid transition temperature at saturated vapour pressure is $T_\mathrm{c}=$\SI{923}{\micro\kelvin}. The sample container is surrounded by two transverse NMR coils that are part of a tank circuit resonator with a $Q$ value of $150$ and a pinch coil used to create an axial minimum of the magnetic field. The resonance frequency of the tank circuit can be tuned in eight equidistant steps between 550~kHz and 833~kHz, corresponding to external magnetic fields between 16.5~mT and 25~mT. The signal is amplified by a cold 
preamplifier\cite{pjheikki_thesis} and room temperature amplifiers. 

The free surface is located 3~mm above the centre of the magnetic field minimum. The location of the free surface is adjusted by removing $^3$He slowly until the desired location is achieved, measuring the pressure of $^3$He gas in a calibrated volume that results from the removal of liquid from the originally fully filled sample container. The outcome is favourably compared with the observed magnon spectrum and a numerical model of the trap. The resulting two traps for magnons are detailed in the next Section.

The time crystal wave function can be written as $\Psi=a e^{-i \omega  t}$, where $t$ is time, $\omega$ is the precession frequency related to the chemical potential $\mu= \hbar \omega$, the phase term $e^{i\varphi}$ is contained in $a$, and the number of magnons $N=|a|^2$. The tipping angle of the precessing magnetisation $\beta_\mathbf{M}$, measured from the magnetic field $\mathbf{H}$, parametrises the spatial profile of the wave function, $N=|a|^2 \propto \int \sin^2 \frac{\beta_\mathbf{M}}{2}\mathrm{d}V$. The signal induced in the pick-up coils (Extended Data Fig.~\ref{Fig:experiment}) is sinusoidal, corresponding to the real part of the rotating complex wave function, $e^{-i \omega t}$. The measured signal amplitude is proportional to the amplitude of the time crystal wave function,

\begin{equation}\label{signal}
 A=c \sqrt{N},
\end{equation}
where $c$ contains the so-called filling factor of the state within the NMR coils, the amplification provided by the tank circuit resonator and other amplifiers the measurement circuit \cite{pjheikki_thesis}, and physical constants \cite{autti2018,zavjalov2015measurements}.

A desired level in the trap can be populated by a radio-frequency pulse via the pick-up coils, followed by slow population decay owing to two mechanisms: The fermionic thermal excitations of the superfluid cause non-hydrodynamic spin diffusion\cite{magnon_relax}. This contribution can be made exponentially small in the zero-temperature limit (1000~s life time has been achieved\cite{2000_ppd}) or dominant at higher temperatures. Observing and controlling the quasi-perpetual time crystal motion inevitably causes also external dissipation,\cite{PhysRevB.101.020505} in our case radiation losses in the measurement circuitry\cite{magnon_relax}. Both of these dissipation mechanisms cause exponential population decay in time, in combination described by time constant $\tau$. The time crystals are well defined provided the life time, which here is $\tau \sim10~$s, is much longer than the time it takes for the time crystal to form after the pulse (here $\tau_E\sim~0.1$s)\cite{PhysRevLett.120.215301,autti2020}. 

The two level system, in the absence of coupling between the states, is described by the wave function $\Psi= b\, e^{-i  \omega_\mathrm{B}  t} + s \,e^{-i \omega_\mathrm{S} t} $, where $b=\sqrt{N_ \mathrm{B}}e^{-i\varphi_\mathrm{B}}$ and $s=\sqrt{N_\mathrm{S}}e^{-i\varphi_\mathrm{S}}$. Only the relative phase enters the dynamics of the system. Hence, $b$ can be chosen to be real, and the combination $b$, $s$ conveniently illustrated by a macroscopic Bloch sphere (Fig.~\ref{Fig:coupling_schematic}c): The surface corresponds to states with total magnon number $N_0=|b|^2 + |s|^2 = N_\mathrm{B} + N_\mathrm{S}$, and the interior to smaller magnon numbers reached during the population decay. The population distribution is given by the polar angle $\theta$ with $N_\mathrm{B}=N_0 \cos(\theta/2)$ and $N_\mathrm{S}=N_0 \sin(\theta/2)$. The relative phase $\phi$ corresponds to the azimuthal angle in the $x$-$y$ plane of the sphere. It evolves in time according to
\begin{equation}\label{rel_phase}
 \phi=\varphi_\mathrm{B} -\varphi_\mathrm{S}+ \int_0^t(\omega_\mathrm{B}-\omega_\mathrm{S})\mathrm{d}t.
\end{equation}
We note that controlling the relative phase is beyond the scope of the present work and requires adjusting the coil geometry. 

\subsection*{Level dynamics in a flexible trap}

$^3$He-B is a p-wave superfluid, hence, the orbital momentum of the Cooper pairs equal to one. In the sample container cylinder, the average orbital momentum $\mathbf{L}$ is distributed symmetrically (``texture'', Extended Data Fig.~\ref{Fig:experiment}) owing to the orienting effects of the magnetic field and the container walls. In addition, we create an axial minimum of $\mathbf{H}$ using a pinch coil, which confines the magnons due to the Zeeman energy. The bulk trapping potential $U({\bf r})=U_\mathbf{H} + U_{\mathbf L}$ therefore has a magnetic part,
\begin{equation}
U_\mathbf{H} =  \hbar\omega_0({\bf r}) \,,
\label{U}
\end{equation}
and a component created by the $\mathbf{L}$ distribution owing to the spin-orbit interaction
\begin{equation}
 U_{\mathbf L}=\hbar \frac{4 \Omega_B^2}{5  \omega_0}\sin^2(\beta_\mathbf{L}({\bf r})/2) \, .
\end{equation}
Here $\omega_0({\bf r})=\gamma |\mathbf{H}|$ is the local Larmor frequency which depends on position ${\bf r}$, $\Omega_B$ is the B-phase Leggett frequency, $\gamma \approx 200~ \mathrm{kHz}\,\mathrm{mT^{-1}}$ the absolute value of the gyromagnetic ratio of $^3$He, and the order parameter distribution is parametrised by the tipping angle of the orbital anisotropy axis, $\beta_L({\bf r})$, measured from the direction of the magnetic field $\mathbf{H}$, oriented along the cylinder axis. 

Bringing the free surface above the trap centre distorts the order-parameter trap as $\beta_\mathbf{L}=0$ at the free surface, creating a local minimum at the surface. Note that we study the time crystals in a frame rotating at the Larmor frequency $\omega_0$ where the uniform magnetic field is absent. Where the notation $\omega_0$ is used without an explicit reference to position, this means Larmor frequency in the middle of the bulk trap, corresponding to the minimum of the harmonic trapping potential. Analysis of the whole observable spectrum in the two traps will be published separately\cite{pjheikki_thesis}. The time crystals located in the two traps can be identified and their frequencies adjusted by changing the profile of the field minimum, aided by the different relaxation rates. Below we concentrate on studying the feedback created by the flexible bulk trap.

The harmonic bulk trap has a radial trapping frequency $\omega_r/(2 \pi) \sim 200$~Hz corresponding to $U_{\mathbf L}$ and an axial trapping frequency $\omega_z/(2 \pi) \sim 20$~Hz corresponding to $U_\mathbf{H}$. The resulting precession frequency is $\omega_0 + \omega_r +\omega_z /2$. Therefore the axial trap can be neglected in the below analysis. A more detailed analysis of the bulk trap can be found in Refs.~\citenum{Heikkinen2014,magnon_relax,zavjalov2015measurements}.

The textural part of the trapping potential feels local magnon density due to spin-orbit interaction: The equilibrium texture minimises a range of free-energy contributions, including the orienting effects of the magnetic field and the sample container walls\cite{thuneberg_texture}. An important additional contribution is the spin-orbit interaction energy
\begin{equation}\label{texture}
F_{\rm so}=  |\Psi({\bf r})|^2 U_{\mathbf L}\,,
\end{equation}
which gives rise to the feedback effect. That is, the bulk trap profile and the shape of the time crystal wave function depend on $N_\mathrm{B}$ so that $\mathrm{d}\omega_\mathrm{B} /\mathrm{d}N_\mathrm{B} < 0$. In the limit of large magnon number the bulk trapping frequency follows \cite{Autti2012}

\begin{equation}\label{fr_dep}
 \omega_\mathrm{B}(N_\mathrm{B})=\bar{\omega}_\mathrm{B}(1-k N_\mathrm{B}^p)
\end{equation}
Here $k>0$ depends on the rigidity of the textural trap and the profile of the magnetic field minimum, $p\approx 5/7$~\cite{Autti2012}, and $\bar{\omega}_\mathrm{B}$ stands for the time crystal trapping frequency in the limit of zero magnons. We emphasise that although $\omega_\mathrm{B}$ changes during the decay of the magnon time crystal, the change is very slow as compared with $\omega_0/(2 \pi)\sim 1~$MHz, and we can thus assume that the wave functions always correspond to the instantaneous trap shape\cite{autti2018}. Note that the surface trap is rigidified by the adjacency of the free surface, and $\omega_\mathrm{S}$ is therefore independent of $N_\mathrm{S}$ to a good approximation.

It is possible to describe the self-trapping effect numerically in a self-consistent calculation of the order parameter texture\cite{thuneberg_texture,kopu_texture}, the resulting trap\cite{zavjalov2015measurements}, the time crystal wave function\cite{Autti2012,autti2012b,autti2018,sautti_thesis}, and population decay \cite{Heikkinen2014,magnon_relax}. That is however not necessary for understanding the experiments presented in this Article, because finding a general form of Eq (\ref{fr_dep}) can be circumvented by fitting and numerical differentiation of the experimental data where necessary, and all other effects can be measured independently. For simplicity, we refer to Eq.~(\ref{fr_dep}) in the below discussion, but the reader should bear in mind that the general form of the nonlinearity is more complicated.

\subsection*{Josephson coupling}

Let us study the observable consequences of the population oscillation. We use the language of the Josephson effect, analogous to the AC Josephson effect\cite{autti2020}, as the oscillation amplitude can only be reliably extracted from the experiment far from the avoided crossing. Near the avoided crossing one should use the more general Rabi oscillation picture.

The amplitude of the AC Josephson population oscillation is 

\begin{equation}\label{Josph_eq}
 \Delta N_\mathrm{B} = \frac{\Omega \sqrt{N_\mathrm{B} N_\mathrm{S}}}{|\omega_\mathrm{B} -\omega_\mathrm{S}|}.
\end{equation}
Here $\Omega$ is the coupling, and $\Delta N_\mathrm{B} \equiv - \Delta N_\mathrm{S}$. The Josephson frequency is $\omega_\mathrm{J} = |\omega_\mathrm{B} - \omega_\mathrm{S}|$. This oscillation modulates the bulk condensate frequency $\omega_\mathrm{B}$ as follows from the self-trapping Eq.~(\ref{fr_dep}). The frequency modulation (FM) is sinusoidal to a good approximation. This is because the amplitude of the population oscillation is small as compared with the total population, and Eq.~(\ref{fr_dep}) can be linearised. 

The resulting instantaneous bulk time crystal frequency $\tilde{\omega}_\mathrm{B}$ can be written as
\begin{equation}
 \tilde{\omega}_\mathrm{B}(t) = \omega_\mathrm{B}(N_\mathrm{B}) + \Delta \omega_\mathrm{B} \cos[(\omega_\mathrm{B}(N_\mathrm{B})-\omega_\mathrm{S}) t].
\end{equation}
Here $\Delta \omega_\mathrm{B}$ is the FM amplitude. It is connected to the population oscillation amplitude $\Delta N_\mathrm{B}$ by 
\begin{equation}
 \Delta \omega_\mathrm{B} = \Delta N_\mathrm{B}~ \mathrm{d}\omega_\mathrm{B}(N_\mathrm{B})/\mathrm{d}N_\mathrm{B}.
\end{equation}
Fourier decomposition of the resulting frequency-modulated signal yields
\begin{equation}
 A_\mathrm{B}^{(n)}= J_n(\Delta \omega_\mathrm{B} / (\omega_\mathrm{B}-\omega_\mathrm{S})) |e^{-i(\omega_\mathrm{B}+n(\omega_\mathrm{B} -\omega_\mathrm{S})) t}|
\end{equation}
Here $J_n$ is the Bessel of the first kind of order $n$. The bulk main trace corresponds to $n=0$. Combining the above expressions, and denoting the first side band ($|n|=1$) amplitude as $A_\mathrm{SB}$, the coupling term can be linearised and expressed in quantities that can be directly measured:

\begin{equation}\label{coupling_eq}
\Omega =\frac{2 A_\mathrm{SB} (\omega_\mathrm{B}-\omega_\mathrm{S})^2}{A_\mathrm{B}^2 A_\mathrm{S} ~\mathrm{d}\omega_\mathrm{B}(A_\mathrm{B})/\mathrm{d}A_\mathrm{B}^2}
\end{equation}
Here we assumed that the filling factors of the bulk and surface states in Eq.~ (\ref{signal}) are equal and constant. Where the time crystal shapes are changing due to changes in the trap profile, the coupling extracted using the above expression is therefore only approximate. 

The side band of the bulk time crystal is seen in Fourier analysis of the experimental signal (Fig.~\ref{Fig:splitting}d). The coupling extracted from this record using Eq.~(\ref{coupling_eq}) extrapolates to $\Omega/(2 \pi)\approx1$~Hz at the crossing, in good agreement with the fitted simulation value $\Omega/(2 \pi)\approx1.4$~Hz. Note that there should be another side band symmetrically below the bulk trace, but it is covered exactly by the surface trace. 

The surface trap is only weakly modified in similar fashion, yielding no visible side bands in the experiment. That is, the AC Josephson effect in a fully rigid trap results in no side bands owing to complex interference of the two wave functions. This can be confirmed by solving the dynamics of the rigid non-decaying coupled system analytically. We used this result to test the validity of the numerical simulation discussed below. Note that Ref.~\onlinecite{autti2020} misleadingly implies that population oscillations directly cause the side bands even in the absence of nonlinear feedback.

Near the avoided crossing one should use the more general Rabi oscillation picture. Solving for the eigenfrequencies of the Hamiltonian in the Rabi regime yields the Rabi frequency $\omega_\mathrm{R}  = \sqrt{(\omega_\mathrm{B}-\omega_\mathrm{S})^2 + (2 \Omega)^2}$ which deviates from $\omega_\mathrm{J}$ if $\Omega> |\omega_\mathrm{B} -\omega_\mathrm{S}|$. This region is not directly visible in the experiments due to interference effects.

\subsection* {Landau-Zener tunnelling}

Let us consider how the time crystal system behaves in the presence of exponential population dissipation,

\begin{equation}
 N_\alpha(t)= N_\alpha(t=0) e^{-2 t/\tau_\alpha} ,
\end{equation}
where $1/\tau_\alpha$ is the relaxation rate of the measured signal (\ref{signal}), and $\alpha$ is either $\mathrm{B}$ for the bulk or $\mathrm{S}$ for surface. For the surface time crystal this makes little difference other than that the population decays slowly. The bulk time crystal frequency $\omega_\mathrm{B}$ depends on $N_\mathrm{B}$ according to Eq.~($\ref{fr_dep}$), and the frequency therefore increases during the decay. Hence, we have obtained the flexible two-level system described by the Hamiltonian (\ref{Hamiltonian}). 

Let us assume $\bar{\omega}_\mathrm{B} >\omega_\mathrm{S}$, and that $N_\mathrm{B}(t=0)$ is large enough so that $\omega_\mathrm{B}(N_\mathrm{B}(t=0)) <\omega_\mathrm{S}$. Now the frequencies of the surface and bulk time crystals will cross in the eigenbasis where $\Omega=0$. If $\Omega>0$ and $N_\mathrm{B}$ decreases adiabatically, magnons in the bulk trap will smoothly move to the surface trap, remaining in the global ground state in an avoided crossing. The minimum frequency separation of the global ground state and the excited state at the avoided crossing is $2\Omega$, as can be solved from the Hamiltonian. 

If the avoided crossing is passed non-adiabatically, a part of the ground state magnon population moves to the excited state. This phenomenon is known as the Landau-Zener-Stueckelberg-Majorana effect. In our case this means that after the avoided crossing some population remains in the bulk trap, which corresponds to the new excited state in the system. The fraction of population promoted to the excited state is 

\begin{equation}
 \delta n = \exp\left(-\frac{2 \pi \Omega^2}{|\partial_t (\omega_\mathrm{B} - \omega_\mathrm{S})|}\right),
\end{equation}
where $\partial_t$ stands for time derivative taken at the avoided crossing.

\subsection*{Numerical simulation}

The time crystal two-level Hamiltonian can be combined with the slow decay into a pair of equations:

\begin{eqnarray}
 i \partial_t \Psi_\mathrm{B} &= &(\omega_\mathrm{B}(N_\mathrm{B}) -i \,\tau_\mathrm{B}^{-1}) \Psi_\mathrm{B} + \Omega \Psi_\mathrm{S} \\
 i \partial_t \Psi_\mathrm{S} &= &(\omega_\mathrm{S} -i \,\tau_\mathrm{S}^{-1}) \Psi_\mathrm{S} + \Omega \Psi_\mathrm{B}.
\end{eqnarray}
Here the right hand side corresponds to the Hamiltonian (\ref{Hamiltonian}), and $i$ is the complex unit. This pair of equations can be solved numerically. Our main motivation for the numerical simulation is to show that the simple two-level Hamiltonian describes the dynamics of the system exhaustively. The most important test for this picture is the avoided crossing and the related population transfer. Reproducing the Rabi oscillations that result in a frequency-modulation of the time crystal frequencies is a secondary test. 

The initial time crystal wave functions, the time crystal decay rates, and the bulk trap self-trapping power law, Eq.~(\ref{fr_dep}), can be extracted from experimental data independently and used as the parameters of the numerical simulation. The coupling $\Omega$ at the avoided crossing cannot be directly extracted from the experiment, and is used as a fitting parameter. To compare with the measured signal we also need the filling factors $c_\alpha$. They are used as fitting parameters as well. 

We note that to reproduce the experimental signals in general, two additional effects need to be included in the simulation: (i) The surface time crystal frequency depends on the population in the bulk trap and (ii) also on the population in the surface trap; $\omega_\mathrm{S}=\omega_\mathrm{B}(N_\mathrm{S},N_\mathrm{S})$. Effect (i) is due to the widening of the constriction that separates the time crystals (Fig.\ref{Fig:coupling_schematic}). This connection is included in the simulation, and it is seen as a decrease of $\omega_\mathrm{S}$ in Extended Data Fig.~\ref{Fig:splitting3} at $t<1~$s, where bulk population is large and surface population negligible. However, this effect can be safely neglected in the analysis of the Landau-Zener effect in Fig.~\ref{Fig:splitting2}, because the avoided crossing takes place at small $N_\mathrm{B}$. The second dependence (ii) is what produces the frequency-oscillations of the magenta line in Extended Data Fig.~\ref{Fig:splitting3}. Both effects can be extracted from experimental data independently.

\section*{Data availability}
The data that supports the findings of this study are available in Zenodo with the identifier DOI:10.5281/zenodo.xxx .
\section*{Code availability}
The simulation codes and guidance in their use are available from the corresponding author upon reasonable request.

 \clearpage

\setcounter{figure}{0}
\renewcommand{\theequation}{S\arabic{equation}} 
\setcounter{equation}{0}
\section*{Extended Data}

 \begin{figure}[hbt!]
 \centerline{\includegraphics[width=1.0\linewidth]{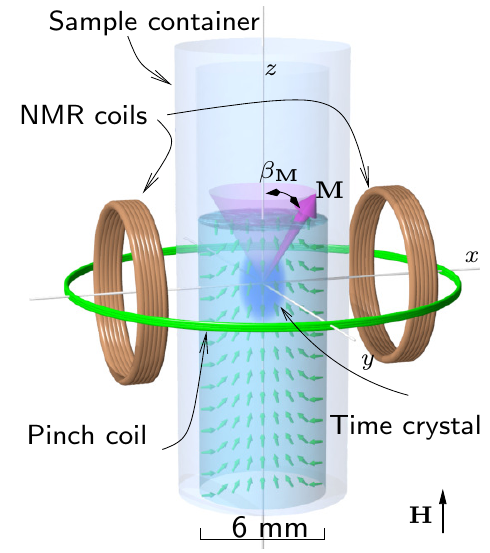}}
   \caption{\label{Fig:experiment} {\bf Schematic illustration of the experiment:} The superfluid $^3$He sample is contained in a quartz glass cylinder. The magnon time crystal (blue blob) is trapped in the middle of the container by the combined effect of a minimum in the static magnetic field, created using a pinch coil (green wire loop), and by the spatial distribution of the superfluid orbital momentum $\mathbf{L}$ (small green arrows). The coherent precession of magnetisation $\mathbf{M}$ (magenta cone) in the time crystal is observed using transverse pick-up coils. The static magnetic field $\mathbf{H}$ is oriented parallel to the axis of the cylinder. The ripple on the superfluid free surface is added for illustrational purposes. The two-level time crystal is schematically illustrated in Fig.~\ref{Fig:coupling_schematic}.}
 \end{figure}

 \begin{figure}[hbt!]
 \centerline{\includegraphics[width=1.0\linewidth]{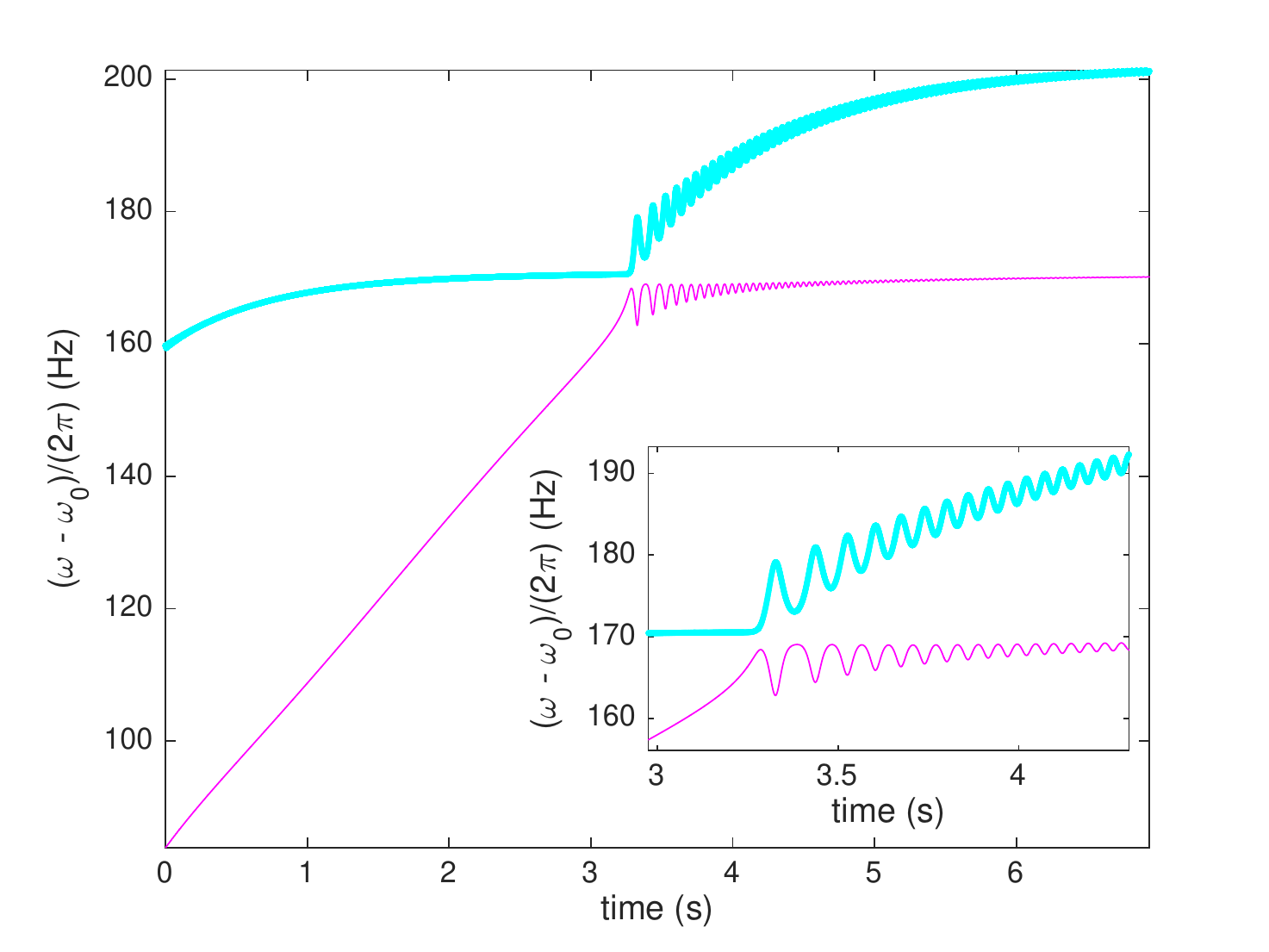}}
   \caption{\label{Fig:splitting3} {\bf Avoided crossing in the simulation:} The dressed ground level time crystal frequency (thin magenta line), extracted from the simulation in Fig.~\ref{Fig:splitting2}, initially corresponds to $\omega_\mathrm{B}$. After the avoided crossing at $t=3.3$~s, it corresponds to $\omega_\mathrm{S}$. Similarly, the excited level frequency (thick cyan line) is initially equal to $\omega_\mathrm{S}$ and switches to $\omega_\mathrm{B}$. In the Rabi regime $\Omega> |\omega_\mathrm{B}-\omega_\mathrm{S}|$ the smallest separation of the dressed frequencies is $2\Omega$. When $N_\mathrm{B}$ is large at $t<1$~s, both $\omega_\mathrm{B}$ and  $\omega_\mathrm{S}$ are reduced owing to the bulk trap expansion. Oscillations of the frequencies after the crossing arise owing to to the population oscillations. The inset shows the avoided crossing in closer detail.
   }
 \end{figure}

\end{document}